\documentclass[a4paper,11pt]{article}
\usepackage{pos}
\usepackage{sidecap}
\sidecaptionvpos{figure}{c}
\usepackage{mciteplus}

\title{Untangling the heavy-flavor mess: status of the Fermilab-MILC calculation of the $B_{(s)}\to D^{(\ast)}_{(s)}\ell\nu$ form factors}
\ShortTitle{Untangling the heavy-flavor mess}

\author*[a,b]{Alejandro Vaquero}
\author[c]{Carleton DeTar}
\author[d,e]{Aida El-Khadra}
\author[f]{Elvira Gámiz}
\author[g]{Steve Gottlieb}
\author[h]{William Jay}
\author[g]{Hwancheol Jeong}
\author[i]{Andreas S. Kronfeld}
\author[d,e]{Andrew Lytle}

\affiliation[a]{Departmento de Física Teórica, Universidad de Zaragoza, 50009 Zaragoza, Spain}
\affiliation[b]{Centro de Astropartículas y Física de Altas Energías (CAPA), 50009 Zaragoza, Spain}
\affiliation[c]{Department of Physics and Astronomy, University of Utah, Salt Lake City, UT 84112-0830, USA}
\affiliation[d]{Department of Physics, University of Illinois Urbana-Champaign, Urbana, IL 61801-3080, USA}
\affiliation[e]{Illinois Center for Advanced Studies of the Universe, University of Illinois Urbana-Champaign, Urbana, IL 61801-3080, USA}
\affiliation[f]{Departamento de Física Teórica y del Cosmos, Universidad de Granada, 18071 Granada, Spain}
\affiliation[g]{Department of Physics, Indiana University, Bloomington, Indiana 47405, USA}
\affiliation[h]{Department of Physics, Colorado State University, Fort Collins, Colorado 80523, USA}
\affiliation[i]{Theory Division, Fermi National Accelerator Laboratory, Batavia, IL 60510-5011, USA}
\affiliation{\textbf{\textsf{Fermilab Lattice and MILC Collaborations}}}

\emailAdd{alexv@unizar.es}

\abstract{\vspace*{-2mm}
We present the status of calculations of the form factors of the most relevant heavy-to-heavy and heavy-to-light decay channels.
Using seven $N_f=2+1+1$ HISQ ensembles, with lattice spacings ranging from 0.15 fm down to 0.06 fm, we calculate the form factors of the decays, including correlations among them.
More than half of our ensembles feature physical pion masses, and the heavy quarks are simulated at their physical masses using the 
Wilson-clover action with the Fermilab interpretation.

Even though we have recently seen huge qualitative and quantitative leaps in the characterization of heavy-to-heavy decays, these advances have failed to translate into improvements for the  inclusive vs exclusive question, or the matter of the Lepton Flavor Universality ratios.
In particular, in the $B\to D^\ast\ell\nu$ channel, the current situation of the lattice-QCD form factors is far from clear.
Further, the latest lattice-QCD results on the heavy-to-light form factors display unexplained tensions that must urgently be resolved.
The work presented here is an attempt to address these issues.}

\FullConference{The 42nd International Symposium on Lattice Field Theory (LATTICE2025)\\
2-8 November 2025\\
Tata Institute of Fundamental Research, Mumbai, India\\}

\renewcommand{\hookAfterAbstract}{%
\par\bigskip
\textsc{Fermilab preprint number}: FERMILAB-CONF-26-0142-T
}

\begin{document}
\maketitle

\section{Introduction}
Nowadays there is a race to find physics beyond the Standard Model (BSM) in the Intensity Frontier, simply because it has become one of the most promising venues:
indirect searches can be several orders of magnitude more sensitive to new physics at large scales, and we are already hitting the limits of our particle accelerators in direct searches.
On the other hand, the flavor sector of the Standard Model (SM) is quite rich in phenomena that may be the subject of new physics searches.
In particular, the $B$~anomalies---understanding here anomaly as a tension between SM calculations and experimental measurements---are an important source of candidates where we expect to find new physics.

\begin{figure}[b]
\centering
\includegraphics[width=0.39\textwidth,angle=0]{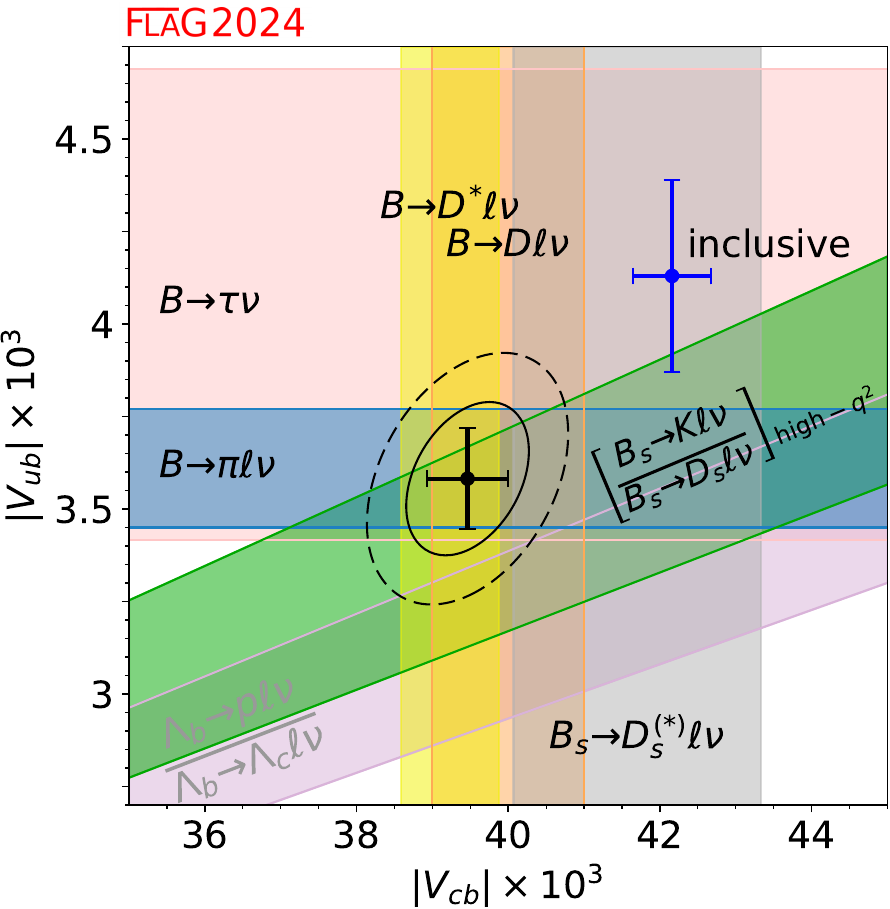}
\includegraphics[width=0.60\textwidth,angle=0]{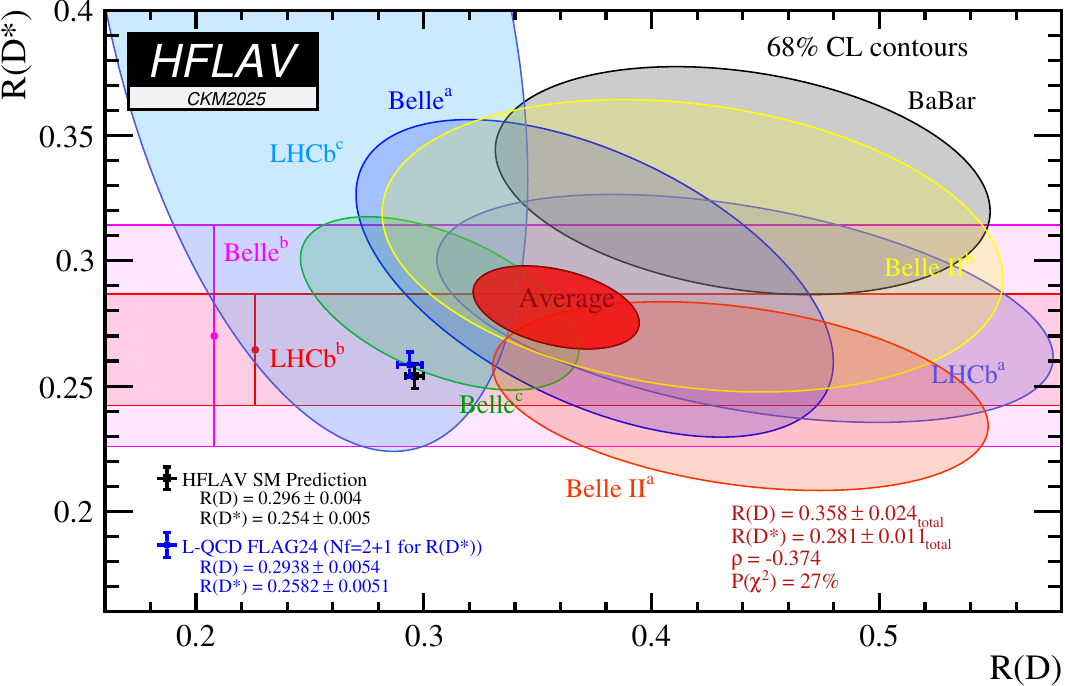}
\caption{Tensions in $|V_{xb}|$ (left) and in $R(D^{(\ast)})$ (right), according to FLAG 2024~\cite{FlavourLatticeAveragingGroupFLAG:2024oxs} and HFLAV~\cite{HFLAV:2024ctg,Belle-II:2024ami}.}
\label{FLAG-Vxb}
\end{figure}

\begin{figure}
\centering
\includegraphics[width=\textwidth,angle=0]{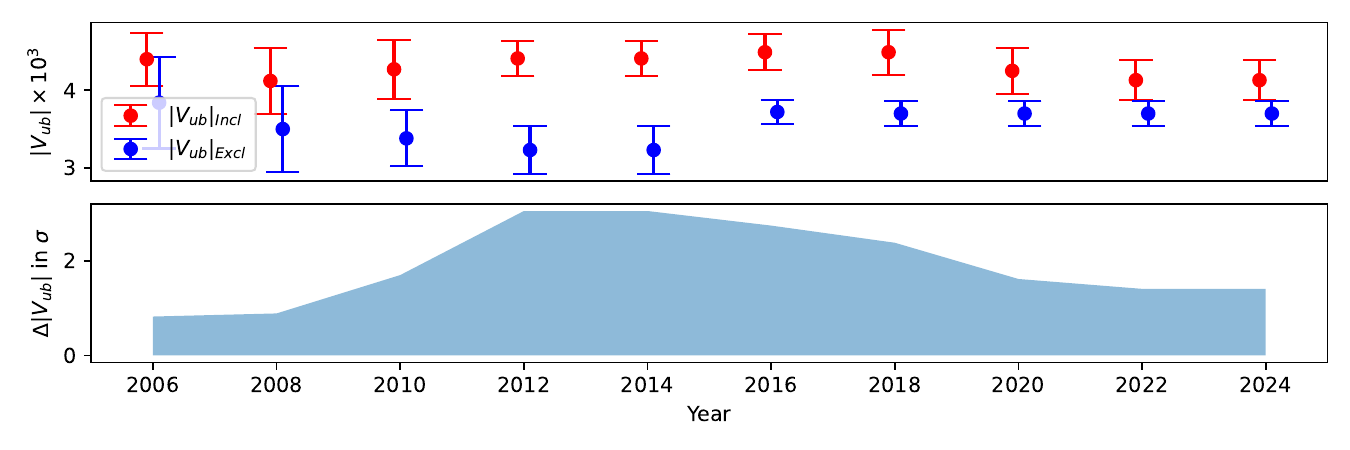}
\caption{{\bf Upper panel:} Evolution of the tension between inclusive and exclusive $|V_{ub}|$ as a function of the year, from the PDG reports.
         {\bf Lower panel:} Difference in $\sigma$ between the inclusive and the exclusive results.}
\label{VubDiff}
\end{figure}

\begin{figure}
\centering
\includegraphics[width=\textwidth,angle=0]{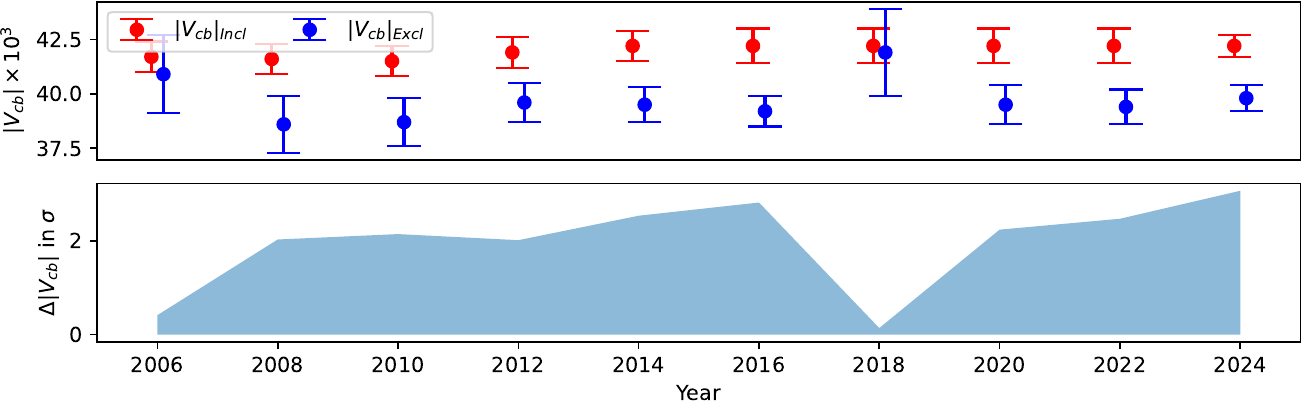}
\caption{{\bf Upper panel:} Evolution of the tension between inclusive and exclusive $|V_{cb}|$ as a function of the year, from the 
PDG reports.
         {\bf Lower panel:} Difference in $\sigma$ between the inclusive and the exclusive results.}
\label{VcbDiff}
\end{figure}

Some of the anomalies that have been the object of long discussions during the last decades arise from the $V_{xb}$ CKM matrix 
elements and the lepton flavor universality (LFU) ratios, both shown in fig.~\ref{FLAG-Vxb}.
%
The left pane shows tension between the inclusive and the exclusive determinations of the CKM matrix elements.
This tension, however, is not split equally between $V_{ub}$ and $V_{cb}$, and the situation of both matrix elements is quite different.
Figure~\ref{VubDiff} shows the evolution of the inclusive-exclusive tensions in $V_{ub}$ over the years, according to the PDG~\cite{ParticleDataGroup:2006fqo,*ParticleDataGroup:2008zun,*ParticleDataGroup:2010dbb,*ParticleDataGroup:2012pjm,*ParticleDataGroup:2014cgo,*ParticleDataGroup:2016lqr,*ParticleDataGroup:2018ovx,*ParticleDataGroup:2020ssz,*ParticleDataGroup:2022pth,*ParticleDataGroup:2024cfk}.
In 2006 there were no tensions, but the errors were large.
As experiments and theoretical calculations improved, the tensions started to increase, but at around 2014 the results began to converge, until the current situation, where the difference is under $2\sigma$.
In contrast, fig.~\ref{VcbDiff} shows the results for $V_{cb}$.
Here a mild tension of about $2\sigma$ or $3\sigma$ has stayed since as early as 2008, and except for the dip in 
2018~\cite{Bigi:2017njr,Grinstein:2017nlq} (which is now understood better),
efforts to either increase this tension to the level of discovery, or reduce it to the level of agreement have not succeeded.
This does not mean that the scientific community has stayed idle. In fact, several lattice-QCD calculations of the form factors of the $B\to D^\ast\ell\nu$ decay have recently been published~\cite{FermilabLattice:2021cdg,Aoki:2023qpa,Harrison:2023dzh},
and these data have resulted in a small reduction of errors in the exclusive determination.
On the other hand, the last years have also brought improvements in the inclusive calculation~\cite{Bordone:2021oof,Bernlochner:2022ucr,Finauri:2023kte}.
Unfortunately, all these advancements have not been enough to elucidate the question of $V_{cb}$.

The other interesting place to look for new physics are the LFU ratios, shown in the right pane of fig.~\ref{FLAG-Vxb}.
A new feature of this plot is an independent point from lattice QCD, taken from the most recent FLAG report~\cite{FlavourLatticeAveragingGroupFLAG:2024oxs}, which agrees very well with previous theoretical expectations.
This point has been made possible due to the most recent lattice-QCD calculations of $B\to D^\ast\ell\nu$~\cite{FermilabLattice:2021cdg,Aoki:2023qpa,Harrison:2023dzh}.
The tensions with experiment, however, are at the level of $\approx 3.5\sigma$, and come mainly from the $R(D)$ contribution.
Even though this is a sizable tension, it is still under the discovery threshold.
Thus, it is urgent to improve our existing results to find out the fate of these $B$~anomalies.

\section{The heavy-to-heavy mess}
\begin{figure}
\centering
\includegraphics[width=0.455\textwidth,angle=0]{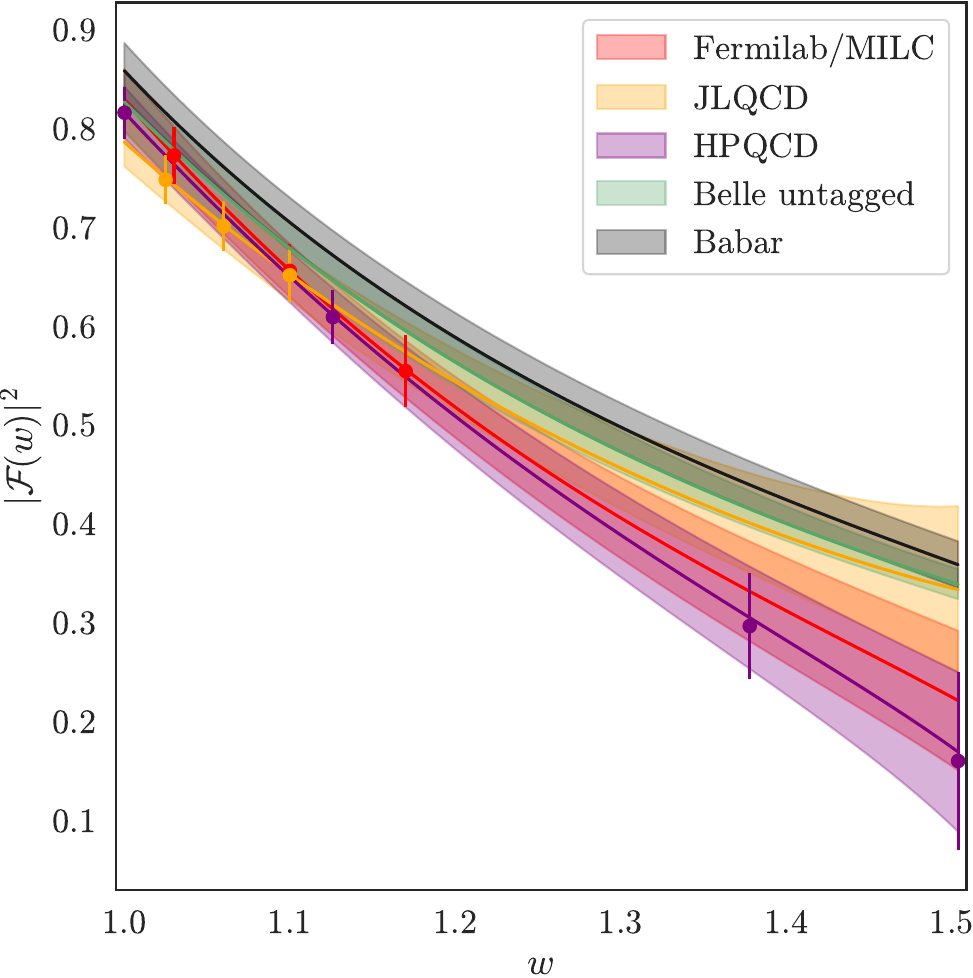}
\includegraphics[width=0.535\textwidth,angle=0]{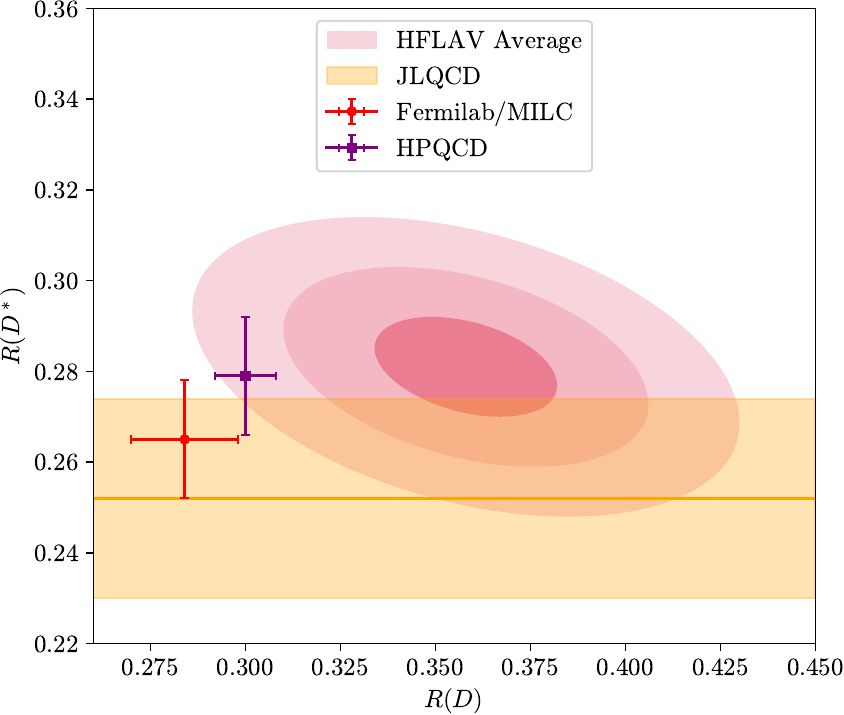}
\caption{Status of the current lattice-QCD calculations, compared against Belle and BaBar experimental measurements.
         {\bf Left:} Decay amplitude extracted from the lattice-QCD form factors and from experiments.
         The $|V_{cb}|$ values used to scale the experimental data come from the respective experiments.
         {\bf Right:} Individual lattice-QCD results for $R(D)$ vs $R(D^\ast)$ compared with the HFLAV average.}
\label{LQCD-BtoDst}
\end{figure}

One of the most exciting developments related to the $B$~anomalies that has taken place in the theory front is the publication of 
several lattice-QCD calculations of the $B\to D^\ast\ell\nu$ form factors.
This channel is gold plated from the experimental point of view, and the $B$~factories have produced a rich amount of data that can be used for precision determinations of $|V_{cb}|$ or $R(D^\ast)$.
What was missing---until very recently---was a precise calculation of the form factors away from the zero recoil point.
The first such result appeared in 2022, by the Fermilab Lattice and MILC collaborations~\cite{FermilabLattice:2021cdg}, soon followed by results from JLQCD~\cite{Aoki:2023qpa} and HPQCD~\cite{Harrison:2023dzh}.
These three calculations use different fermion actions and gauge ensembles, and not only are they completely independent, but also feature different systematic errors.
Thus, one can accurately assess where the state-of-the-art lattice-QCD calculations stand by comparing the three results.
Figure~\ref{LQCD-BtoDst} shows the current results for the decay amplitude of the decay, as well as for the LFU ratios.
The left pane shows the most interesting plot: a comparison between the decay amplitudes of different lattice-QCD calculations, and those obtained in Belle~\cite{Waheed:2018djm} and BaBar~\cite{Dey:2019bgc} experiments.
Some calculations display a mild tension with the experimental curve, and there has been some speculation about what this could mean. 
However, even though the results might look messy, the agreement between different lattice-QCD calculations is relatively good: the differences in $R(D^\ast)$ are below $1\sigma$, and those in the decay amplitude are well under $2\sigma$ in all cases.
One could easily perform combined fits of the lattice data, for instance, a BGL fit with quadratic coefficients for the four $B\to 
D^\ast\ell\nu$ form factors, and obtain good results without tensions, with $p$~values of order~1. 
On the other hand, a combined BGL fit of the BaBar and Belle data sets for the three form factors involved in this decay 
amplitude results in a much higher tension, with a $p$~value of order $10^{-2}$.
The dispersion between the different lattice-QCD results can be understood as a systematic error, and the main criticism one can 
come up with against the lattice-QCD results is that we need to improve our precision.
Hence, it is clear that we need more and better calculations of the form factors of this channel.

\section{The heavy-to-light mess}
The picture is quite different in the case of heavy-to-light decays.
Here the tension in the corresponding CKM matrix element, $V_{ub}$, is not too uncomfortable, but a much more serious issue arises: 
the lattice-QCD calculations simply do not agree very well with each other.
Whereas in the $B\to D^\ast\ell\nu$ channel the form factors show a healthy dispersion, well within compatibility bounds, the 
latest FLAG report highlight serious circumstances with the $B\to\pi\ell\nu$ and $B_s\to K\ell\nu$ form factors~\cite{FlavourLatticeAveragingGroupFLAG:2024oxs}.
\begin{figure}
\centering
\includegraphics[width=0.49\textwidth,angle=0]{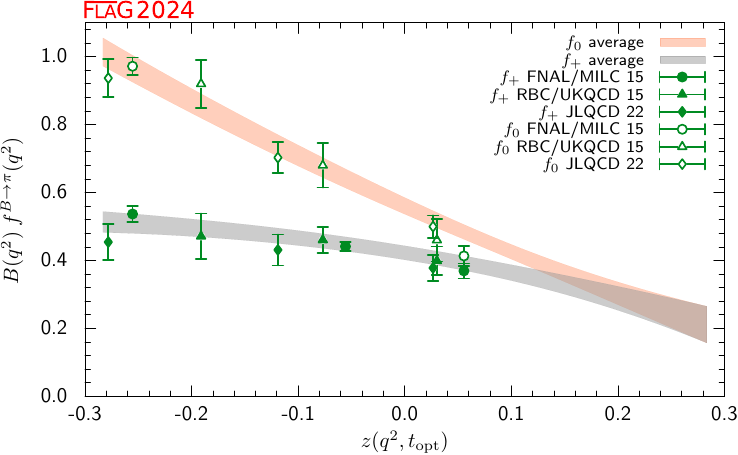}
\includegraphics[width=0.49\textwidth,angle=0]{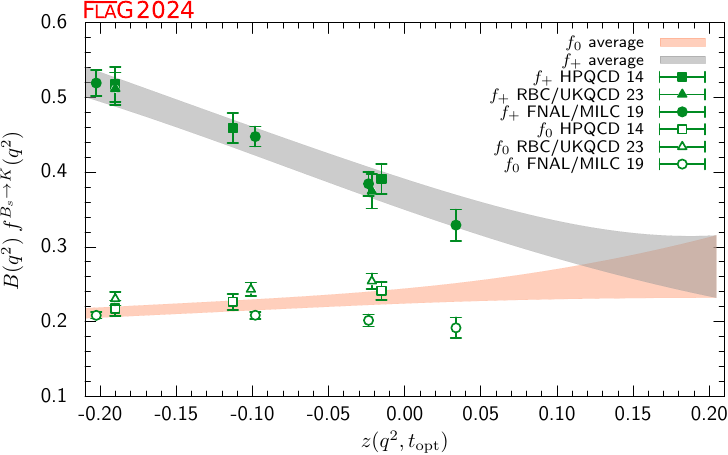}
\caption{Tensions in the $B\to\pi\ell\nu$ (left) and the $B_s\to K\ell\nu$ form factors, according to FLAG 2024.}
\label{LQCD-BtoLight}
\end{figure}
The combined fits yield values of $\chi^2/\textrm{dof} > 3$, signaling an incompatibility of the data produced by different collaborations.
Since FLAG is the resource experimentalist and phenomenologist use to access lattice-QCD results, the inconsistency might result in a lower confidence in the data generated by the lattice community.
It is urgent to find an explanation.

\begin{SCfigure}[1]
\centering
\includegraphics[width=0.4\textwidth,angle=0]{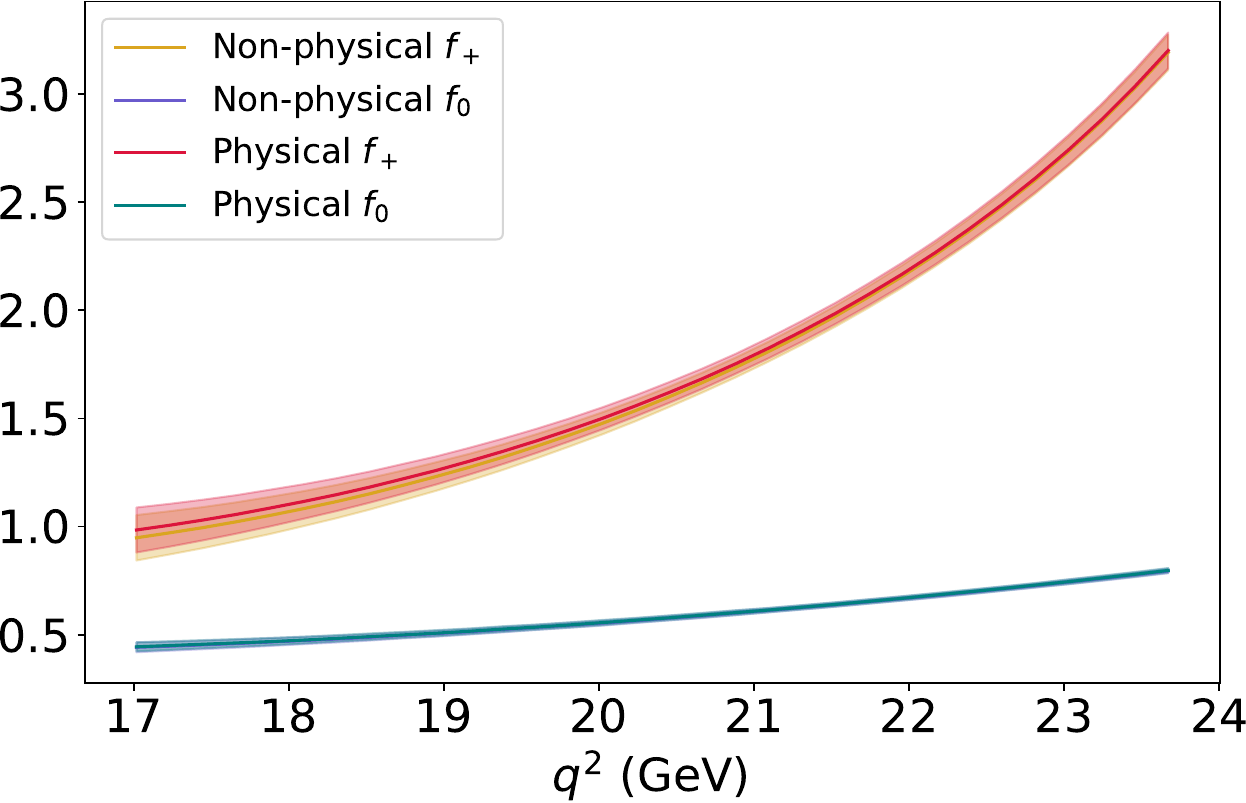}
\caption{Fermilab Lattice and MILC collaborations reanalysis of their $B_s\to K\ell\nu$ form factors. The way the continuum limit is taken does not seem to affect the final results in this case.
         The $B\to\pi\ell\nu$ form factors (not shown) display a similar behavior.}
\label{StillGood}
\end{SCfigure}

A step in the right direction was taken by the RBC/UKQCD collaboration in their latest $B_s\to K\ell\nu$ calculation~\cite{Flynn:2023nhi}.
Normally, in the pseudoscalar to pseudoscalar decays it is more comfortable to work with the form factors $f_\parallel$ and 
$f_\bot$, which are linear combinations of the usual $f_+$ and $f_0$ form factors.
The key difference is that $f_+$ and $f_0$ correspond to a given angular momentum of the lepton-neutrino system, with implications 
stemming from unitarity and analyticity.
One can take the continuum limit in either basis, and the two procedures can give different results, especially for the form factor $f_0$.
Because $f_\parallel$ and $f_\bot$ 
do not correspond to well-defined angular momentum, certain unphysical assumptions about pole 
factors must be made while taking the continuum limit.
These assumptions might be more or less innocuous, depending on the precision of the data and the lattice action, but certainly 
they are a new source of systematic errors that can easily be removed by just reconstructing $f_+$ and $f_0$ before taking the continuum limit.
The Fermilab Lattice and MILC collaborations performed tests in their 2015 and 2019 calculations of the $B\to\pi\ell\nu$ and $B_s\to K\ell\nu$ form factors~\cite{FermilabLattice:2015mwy,FermilabLattice:2019ikx}, to assess to which extent they were affected.
The results of this analysis are shown in fig.~\ref{StillGood}: the differences between both procedures result in compatible results, well within one $\sigma$.
The discrepancy found by RBC/UKQCD is scarcely visible in the $D$ decay analysis of~\cite{FermilabLattice:2022gku}.
Even so, it is not enough to  explain the tensions among different lattice-QCD calculations.

\section{The way forward}
We have laid out reasons why better calculations are needed in both heavy-to-heavy and heavy-to-light sectors.
The Fermilab Lattice and MILC collaborations are committed to improving these results, and for this reason we have two different calculations running simultaneously.
One of them uses the HISQ action~\cite{Follana:2006rc} for the light and the heavy quarks (HISQ on HISQ), and its status has been reported in other talks in this conference~\cite{Chauhan:2025abc}.
This section describes the other calculation.

\subsection{Ensembles and setup}
\begin{SCfigure}[1]
\centering
\includegraphics[width=0.5\textwidth,angle=0]{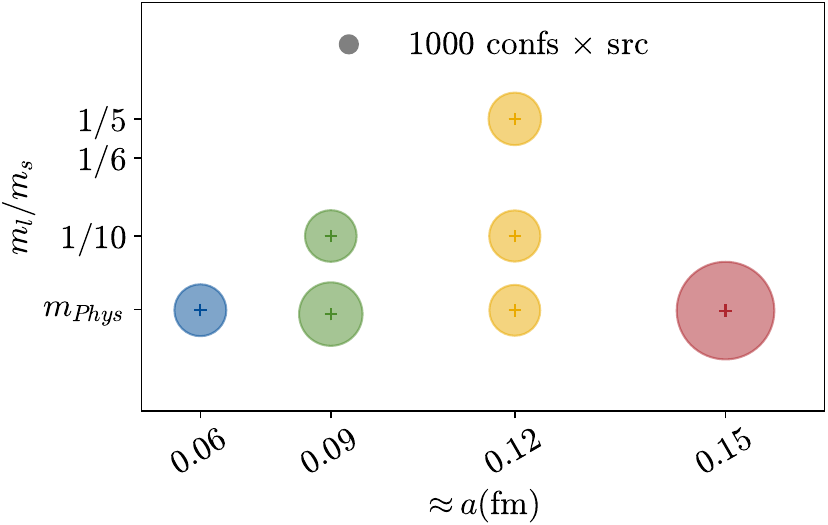}
\caption{Ensemble list of the calculation. The size of each circle is proportional to the available statistics.}
\label{Ensembles}
\end{SCfigure}

This calculation uses seven $N_f=2+1+1$ HISQ ensembles, with four different lattice spacings ranging from $0.15$ fm down to $0.06$ fm.
There is one ensemble per lattice spacing with light quark masses tuned to result in physical pion masses.
The ensemble distribution along with their statistics are show in fig.~\ref{Ensembles}.
The heavy quarks, $b$ and $c$, in the valence sector are simulated using the Fermilab interpretation~\cite{El-Khadra:1996wdx} of the clover action.
This project aims to perform a fully correlated analysis of the four heavy-to-heavy channels, $B_{(s)}\to D^{(\ast)}_{(s)}\ell\nu$, 
and the three heavy-to-light channels, $B\to\pi\ell\nu$, $B_s\to K\ell\nu$ and $B\to K\ell\ell$.
Moreover, this calculation uses three ensembles in common with the HISQ-on-HISQ calculation~\cite{Chauhan:2025abc}, so a 
focused analysis of these three can be used to assess the performance of different heavy-quark actions applied to the same problem.

\subsection{Heavy-to-heavy form factors}
\begin{figure}
\centering
\includegraphics[width=0.45\textwidth,angle=0]{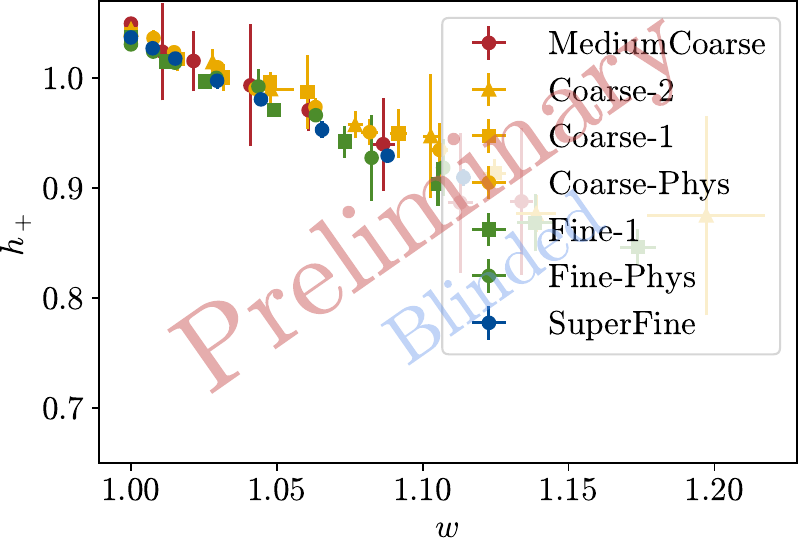}
\includegraphics[width=0.49\textwidth,angle=0]{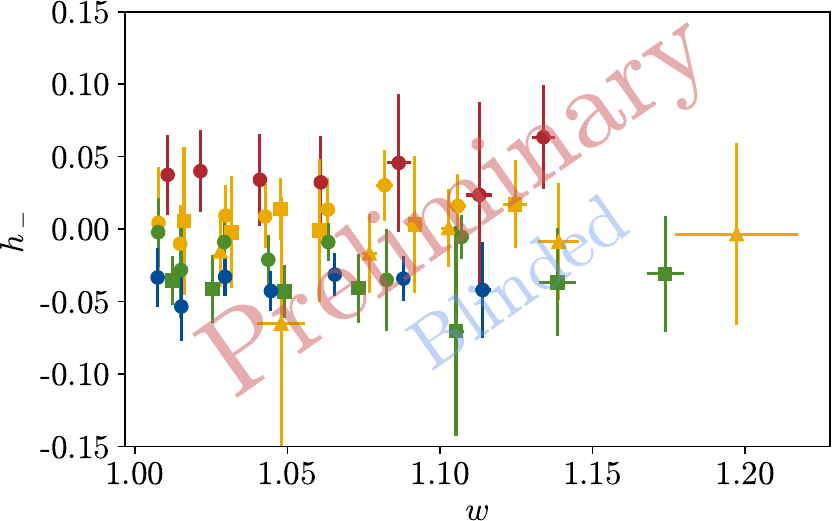}
\caption{Preliminary and blinded results for the $h_{+,-}$ (left, right) form factors for the $B\to D\ell\nu$ decay.}
\label{BtoDFF}
\end{figure}

\begin{figure}
\centering
\includegraphics[width=0.45\textwidth,angle=0]{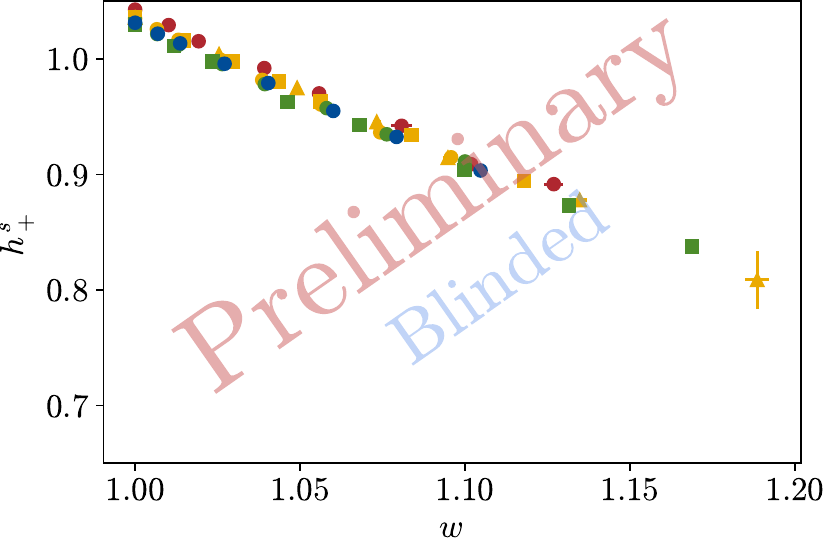}
\includegraphics[width=0.49\textwidth,angle=0]{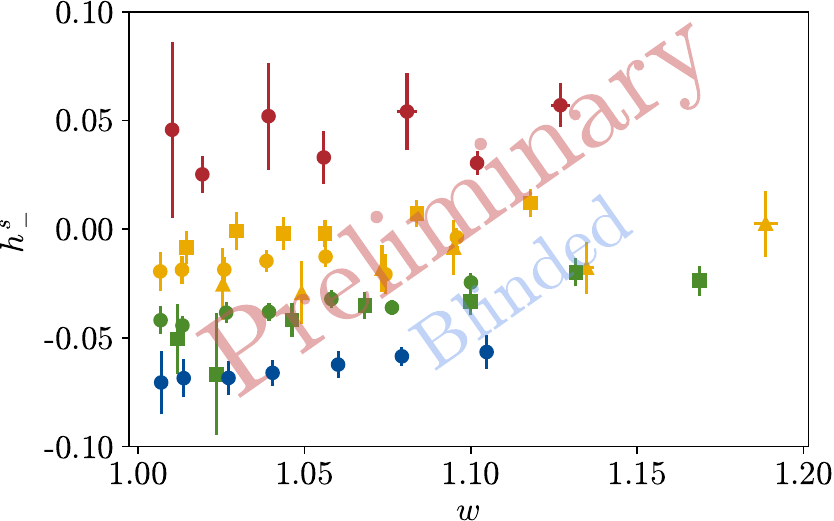}
\caption{Same as fig.~\ref{BtoDFF}, but for the $B_s\to D_s\ell\nu$ channel.}
\label{BstoDsFF}
\end{figure}

\begin{figure}
\centering
\includegraphics[width=0.48\textwidth,angle=0]{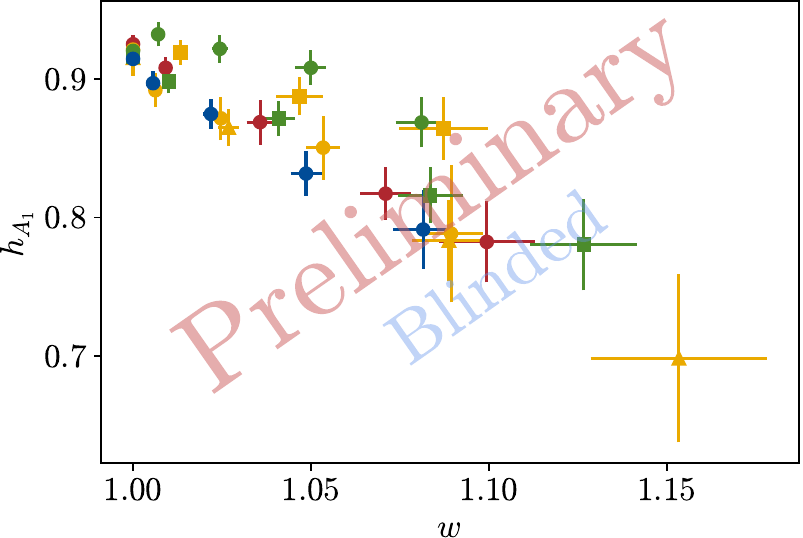}
\includegraphics[width=0.51\textwidth,angle=0]{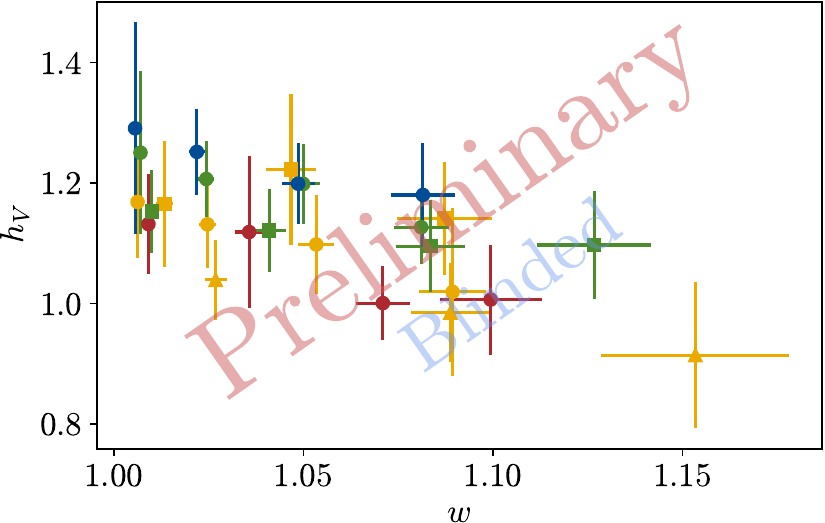}
\caption{Preliminary and blinded results for the $h_{{A_1},V}$ (left, right) form factors for the $B\to D^\ast\ell\nu$ decay.}
\label{BtoDstFF1}
\end{figure}

\begin{figure}
\centering
\includegraphics[width=0.48\textwidth,angle=0]{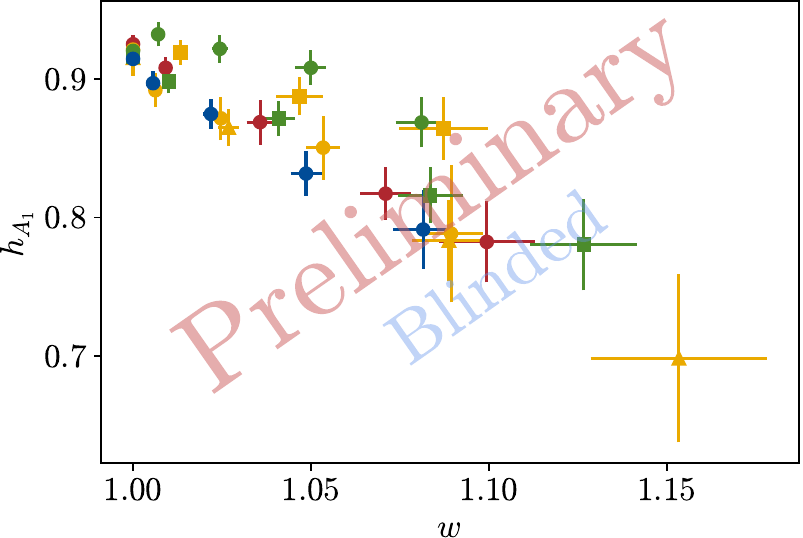}
\includegraphics[width=0.51\textwidth,angle=0]{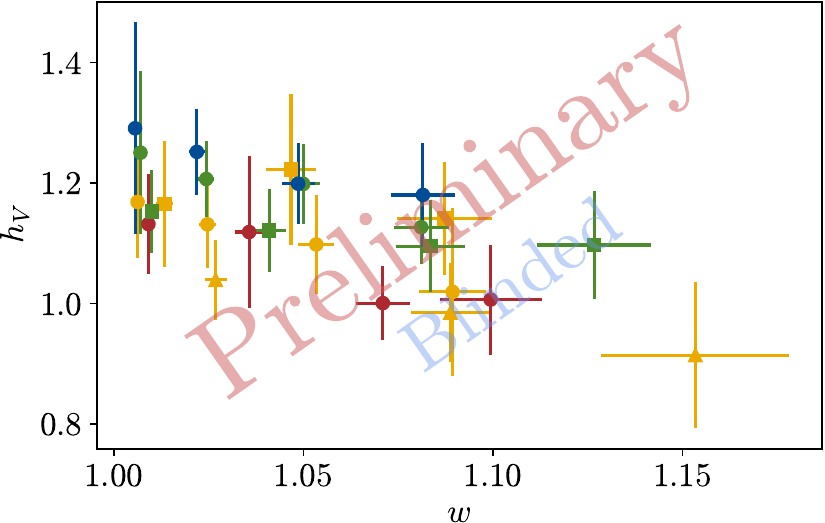}
\caption{Same as fig.~\ref{BtoDstFF1}, but for the $B_s\to D_s^\ast\ell\nu$ channel.}
\label{BstoDsstFF1}
\end{figure}

Currently, we have calculated the form factors for all the heavy-to-heavy channels.
In figs.~\ref{BtoDFF},~\ref{BstoDsFF},~\ref{BtoDstFF1} and~\ref{BstoDsstFF1} we show a selection of form factors for all the channels.
The results, although very preliminary and still blinded, seem consistent with existing calculations of the form factors.
As expected, the form factors for the $B_s$ decays can be calculated with a much higher precision than those for the $B$.
The next steps are to take the continuum limit and perform a careful analysis of the systematic errors.

\subsection{Heavy-to-light form factors}
\begin{figure}
\centering
\includegraphics[width=0.329\textwidth,angle=0]{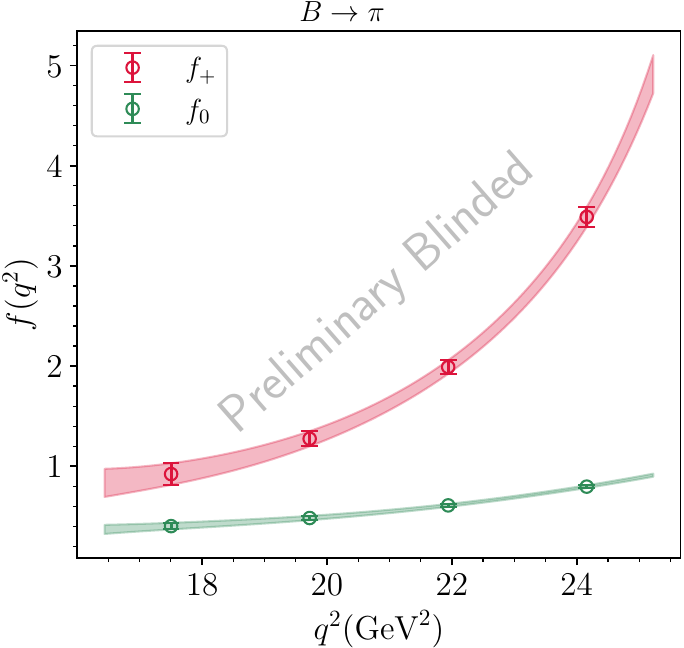}                                                                                                                                                                                                                                                       
\includegraphics[width=0.329\textwidth,angle=0]{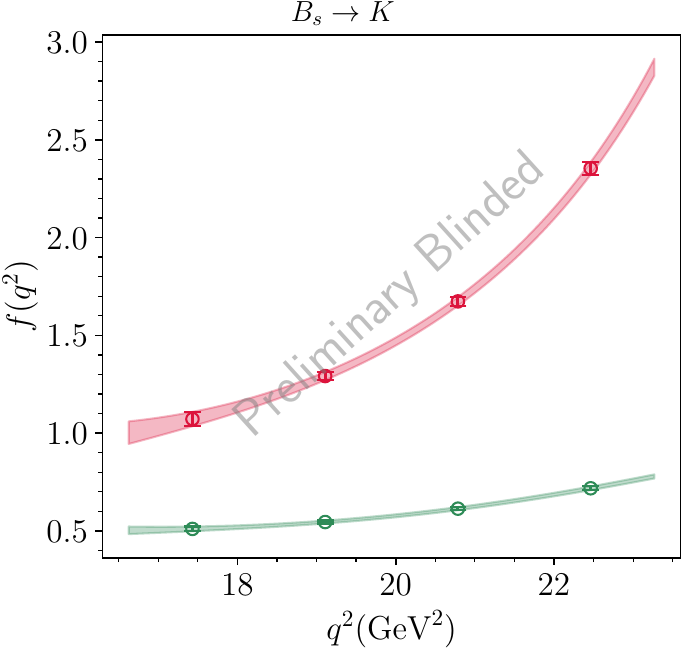}
\includegraphics[width=0.329\textwidth,angle=0]{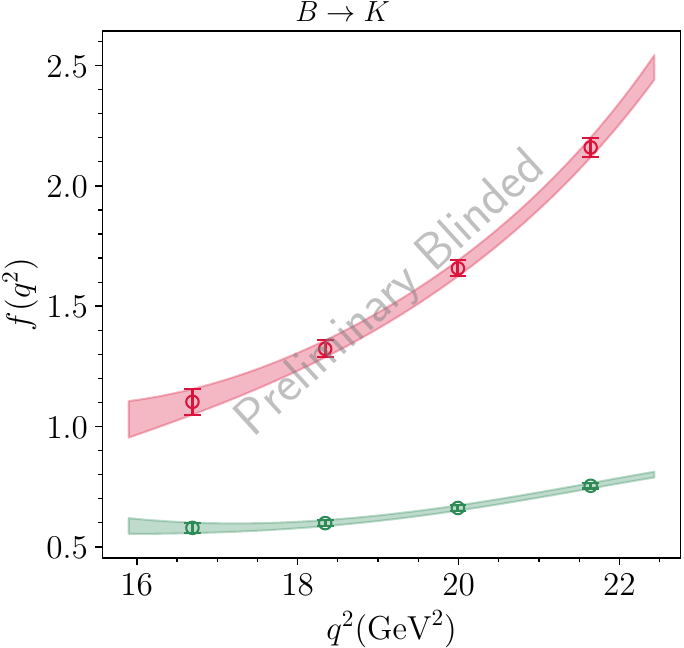}
\caption{From left to right, preliminary and blinded results for the chiral continuum limit of the $f_+$ and $f_0$ form factors of the $B\to\pi\ell\nu$, $B_s\to K\ell\nu$ and the $B\to K\ell\ell$ decays.}
\label{BtoLightFF}
\end{figure}

The analysis of the heavy-to-light channels is in a slightly more advanced state.
In fig.~\ref{BtoLightFF} we show preliminary results for the chiral-continuum extrapolation.
We expect to finalize the analysis in the coming months.

\section{Conclusions}
The current status of the lattice-QCD calculations of the form factors in the heavy-to-heavy and heavy-to-light semileptonic decays is unsatisfactory:
On the one hand, even though the heavy-to-heavy calculations have greatly improved during the latest years, and show a good agreement, they have failed to solve the key issues related to the $B$~anomalies, i.e., $|V_{cb}|$ and the LFU ratios.
With experiments like LHCb and Belle~II generating results at a quick pace, we must match their efforts and improve our results.
On the other hand, the heavy-to-light form factor calculations show discrepancies between lattice-QCD calculations from different 
collaborations. 

Our collaboration has a well-defined roadmap to address these issues, with two different calculations that will deliver high-quality results.
The calculation presented in this work comprises both heavy-to-heavy and heavy-to-light decay channels.
It is in a relatively advanced state, and we expect to deliver results in the coming months.

\acknowledgments

Computations for this work were carried out with resources provided by the USQCD Collaboration; by the ALCF and NERSC, which are 
funded by the U.S. Department of Energy. 
This work was partly supported by the U.S. Department of Energy, Office of Science, under grants No.\ DE-SC0010120 (S.G.), No.\ DE-No.\ DE-SC0021006 (W.J.), No.\ SC0015655 (A.X.K., A.T.L.);
by the U.S. National Science Foundation under Grants Nos.\ PHY20-13064 and PHY23-10571 (A.V., C.D.), and Grant No.\ 2139536 for Characteristic Science Applications for the Leadership Class Computing Facility (H.J.);
by the Simons Foundation under their Simons Fellows in Theoretical Physics program (A.X.K.);
by the Exascale Computing Project (17-SC-20-SC), a collaborative effort of the U.S. Department of Energy, Office of Science, and the National Nuclear Security Administration (H.J.);
by MICIN/AEI/10.13039/501100011033 and FEDER (EU) under Grant PID2022-140440NB-C21 and Junta de Andalucía under Grant No.\ FQM-101 (E.G.);
by Consejeria de Universidad, Investigacion e Innovacion and Gobierno de España and EU–NextGeneration, under Grant AST228.4(E.G.).
This document was prepared by the Fermilab Lattice and MILC Collaborations using the resources of the Fermi National Accelerator Laboratory (Fermilab), a U.S.\ Department of Energy, Office of Science, HEP User Facility.
Fermilab is managed by Fermi Forward Discovery Group, LLC, acting under Contract No.~89243024CSC000002 with the U.S.\ Department of Energy.

\bibliographystyle{JHEP}
\bibliography{PoSLat25}

\end{document}